\pdfoutput=1
\documentclass[english,showkeys]{revtex4-1}
\usepackage[T1]{fontenc}
\usepackage[latin9]{inputenc}
\usepackage{color}
\usepackage{babel}
\usepackage{amstext}
\usepackage{graphicx}
\usepackage[unicode=true,pdfusetitle,
 bookmarks=true,bookmarksnumbered=false,bookmarksopen=false,
 breaklinks=false,pdfborder={0 0 0},pdfborderstyle={},backref=false,colorlinks=true]
 {hyperref}

\makeatletter

\newcommand{\noun}[1]{\textsc{#1}}

\@ifundefined{showcaptionsetup}{}{%
 \PassOptionsToPackage{caption=false}{subfig}}
\usepackage{subfig}
\makeatother

\begin{document}
\title{MFV: Application software for the visualization and characterization
of the DC magnetic field distribution in circular coil systems}
\author{J. D. Alzate-Cardona$^{1}$, D. Sabogal-Suárez$^{1}$, J. Torres$^{2}$,
E. Restrepo-Parra$^{1}$}
\affiliation{$^{1}$PCM Computational Applications, Departamento de Física y Química,
Facultad de Ciencias Exactas y Naturales, Universidad Nacional de
Colombia, Manizales, Colombia~\\
$^{2}$Research Group on Electromagnetic Fields, Environment and Public
Health, Departamento de Física, Facultad de Ciencias Exactas y Naturales,
Universidad de Caldas, Manizales, Colombia}
\begin{abstract}
The characterization of the magnetic field distribution is essential
in experiments and devices that use magnetic field coil systems. We
present an open-source application software, \noun{MFV} (Magnetic
Field Visualizer), for the visualization of the distribution of the
magnetic field produced by circular coil systems. MFV models, simulates,
and plots the magnetic field of coil systems composed by any number
of circular coils of any size placed symmetrically along the same
axis. Therefore, any new design or well known coil system, such as
the Helmholtz or the Maxwell coil, can be easily modeled and simulated
using MFV. A graph of the homogeneity of the magnetic field can be
also produced, showing the work region where the magnetic field is
homogeneous according to a percentage of homogeneity given by the
user. An standardized input and output file format is employed to
facilitate the exchange and archiving of data. We include some results
obtained using MFV, showing its applicability to characterize the
magnetic field in different coil systems. Furthermore, the magnetic
field results provided by MFV were validated by comparing them with
results obtained experimentally in a Helmholtz coil system.
\end{abstract}
\maketitle

\section{Introduction}

The generation of homogeneous magnetic fields is required in a broad
range of applications. For instance, magnetic fields are commonly
used to stimulate biological systems \citep{Malmivuo1995}, particularly
on in vitro experiments, and for the calibration of magnetic field
sensors and probes \citep{Bronaugh}. The accuracy and reliability
of the experiments developed on these kind of applications depend
on different factors, but mainly on the distribution and homogeneity
of the magnetic field. The most common coil system to produce homogeneous
and controlled magnetic fields is the Helmholtz coil; however, its
homogeneous work region is not large enough for some applications.
Therefore, many efforts have been made in order to produce large homogeneous
regions, and several coil systems of different sizes, shapes, and
number of coils have been proposed \citep{Gottardi2003,Maxwell2010,Merritt1983,Raganella1994,Wang2002,Kirschvink1992}. 

In order to employ a coil system for any kind of application that
requires a work region with an homogenous magnetic flux density, it
is essential to characterize the magnetic field distribution, and
use this information to find the size and shape of a potential region
of work where the magnetic field has a specific homogeneity value.
Therefore, the development of tools to model and simulate the spatial
distribution of the magnetic flux density produced by coil systems
can be very useful to carry out experiments in different fields of
study. For this reason, we present an open-source application software,
MFV (Magnetic Field Visualizer), for the modeling, simulation, and
visualization of the magnetic flux density distribution and homogeneity
of circular coil systems. MFV can be used as a pedagogical tool to
improve the understanding of the magnetic field distribution, or it
can be used by an experimentalist to characterize the magnetic field
of a coil system for an experiment or application, to design a new
coil system, or to optimize experimental methodologies. Furthermore,
high quality graphs of the magnetic field distribution and homogeneity
can be generated using MFV.

\section{MFV application software}

The aim of MFV is to provide the tools to easily model, simulate,
and visualize the magnetic field distribution of any kind of circular
coil system. The main features of \noun{MFV} are the following:
\begin{itemize}
\item High flexibility to model and simulate a great variety of magnetic
field coil systems. These systems can be composed by any number of
circular coils of any size placed symmetrically along the same axis.
The number of turns and the electric current through each coil can
also be set by the user.
\item Calculation and visualization of the magnetic flux density distribution
and its homogeneity. The visualization corresponds to that of a plane
parallel to the axis of symmetry of the coil system, which is enough
to characterize the magnetic field distribution in all the system
thanks to its rotational symmetry.
\item The homogeneous work region is calculated and visualized according
to a percentage of homogeneity given by the user. Parameters of a
practical experimentation volume, where the magnetic field is homogeneous,
are given to the user.
\item A group of presets of common coils systems with pre-established parameters
are provided, including the Helmholtz coil, the Maxwell coil \citep{Maxwell2010},
the Tetracoil \citep{Gottardi2003}, the Wang coil \citep{Wang2002},
and the Lee-Whiting coil \citep{Kirschvink1992}. 
\item The user can set different simulation parameters, including the size
of the simulation region and the number of points of the simulation.
\item Availability of options to make zoom and produce graphs of the magnetic
field distribution and homogeneity.
\item Input parameters and simulation results can be saved and loaded into
the application. Furthermore, experimental results formatted accordingly
can be loaded into the application to characterize the magnetic field
using the available tools. The XLSX format is employed to facilitate
the exchange and archiving of data.
\item An estimation of the simulation time is given during the simulation.
\item \noun{MFV} does not require to be built in order to be executed, it
is composed by executables that can be run directly on \noun{Linux
}64 bits, \noun{macOS High Sierra }(or later\noun{),} and \noun{Windows
}8 (or later) operating systems. The executables are available for
download at the MFV GitHub repository \citep{Alzate-Cardona2019}.
\end{itemize}

\subsection{Model and method}

\subsubsection{Magnetic field}

The off-axis magnetic field (magnetic flux density) generated by a
single coil placed at a point $Z$ on the axis of symmetry of the
coil ($z$-axis) is given by \citep{Smythe1939}

\begin{equation}
B_{\rho}(\rho,z)=\frac{\mu_{0}}{2\pi}\frac{NI(z-Z)}{\rho[(r+\rho)^{2}+(z-Z)^{2}]^{1/2}}\times\left[\frac{r^{2}+\rho^{2}+(z-Z)^{2}}{(r-\rho)^{2}+(z-Z)^{2}}K_{2}-K_{1}\right]\label{eq:1}
\end{equation}

and

\begin{equation}
B_{z}(\rho,z)=\frac{\mu_{0}}{2\pi}\frac{NI}{[(r+\rho)^{2}+(z-Z)^{2}]^{1/2}}\times\left[\frac{r^{2}-\rho^{2}-(z-Z)^{2}}{(r-\rho)^{2}+(z-Z)^{2}}K_{2}-K_{1}\right]\text{,}\label{eq:2}
\end{equation}
where $B_{\rho}$ and $B_{z}$ are the $\rho$ and $z$ components
of the magnetic field in cylindrical coordinates, respectively, $\mu_{0}$
is the vacuum permeability, $N$, $r$, and I are the number of turns,
radius, and the electric current through the coil, and $K_{1}$ and
$K_{2}$ are the first and second complete elliptic integrals with
modulus $k$, which is given by

\begin{equation}
k^{2}=\frac{4r\rho}{(r+\rho)^{2}+(z-Z)^{2}}\text{.}
\end{equation}

When a coil system is simulated in MFV, the contribution of each coil
is calculated according to Equations \ref{eq:1} and \ref{eq:2} for
all the points in the simulation region.

\subsubsection{Homogeneity}

The homogeneity is computed with respect to the value of the total
magnetic flux density at the center of the simulation region. It is
calculated as the relative difference between the value of the total
magnetic field at any point ($B(\rho,z)$) and the value at the center
of the simulation region ($B_{o}$), as follows

\begin{equation}
h=1-\frac{B(\rho,z)-B_{o}}{B_{o}}\text{.}
\end{equation}

Therefore, an homogeneity of $98\%$ corresponds to a region where
the values of the magnetic field do not vary more than $2\%$ from
the value of reference.

\subsection{\noun{I}nterface and usage}

MFV is a very intuitive and easy-to-use application software. Several
tooltips and alert messages are included in order to guide the user
during the input of parameters and visualization. The interface is
composed by three main sections, the input parameters window, the
results window, and the zoom and homogeneity windows.

\subsubsection{Input parameters}

Once the MFV executable is run, the input parameters window pops up,
as shown in Figure \ref{fig:MFV-input-parameters}. In this window,
the user inputs the parameters that describe the coil system to be
modeled and simulated. The menu bar located at the top of the window
houses three drop-down menus. The ``File'' menu provides options
to refresh the window (``New''), load input parameters (``Load
parameters''), load simulation results (``Load results''), and
quit the application (``Quit''). Input results and simulation parameters
must be loaded from a file in XLSX format. The ``Presets'' menu
contains a group of common coils systems with pre-established parameters.
And the ``Help'' menu provides an option (``About'') to get information
about the software. Below the menu bar the user can input the parameters
of the coils, including the radius, number of turns, electric current,
and position of the coil along the axis of symmetry ($z$-axis). Buttons
to add (``+'') and remove (``-'') coils are provided. Finally,
at the lower part of the window there is a button (``Simulate'')
to start the simulation of the coil system. Next to the ``Simulate''
button there is a checkbox to decide whether to use default parameters
to carry out the simulation. If the box is unchecked and the ``Simulate''
button is clicked, a window pops up allowing the user to set the simulation
limits (in cartesian coordinates) and the number of points, as shown
in Figure \ref{fig:simulation-parameters}. When the simulation is
running, a small window showing the progress and remaining time of
the simulation is displayed on the screen.
\begin{center}
\begin{figure}[h]
\begin{centering}
\subfloat[\label{fig:MFV-input-parameters}]{\begin{centering}
\includegraphics[width=0.5\paperwidth]{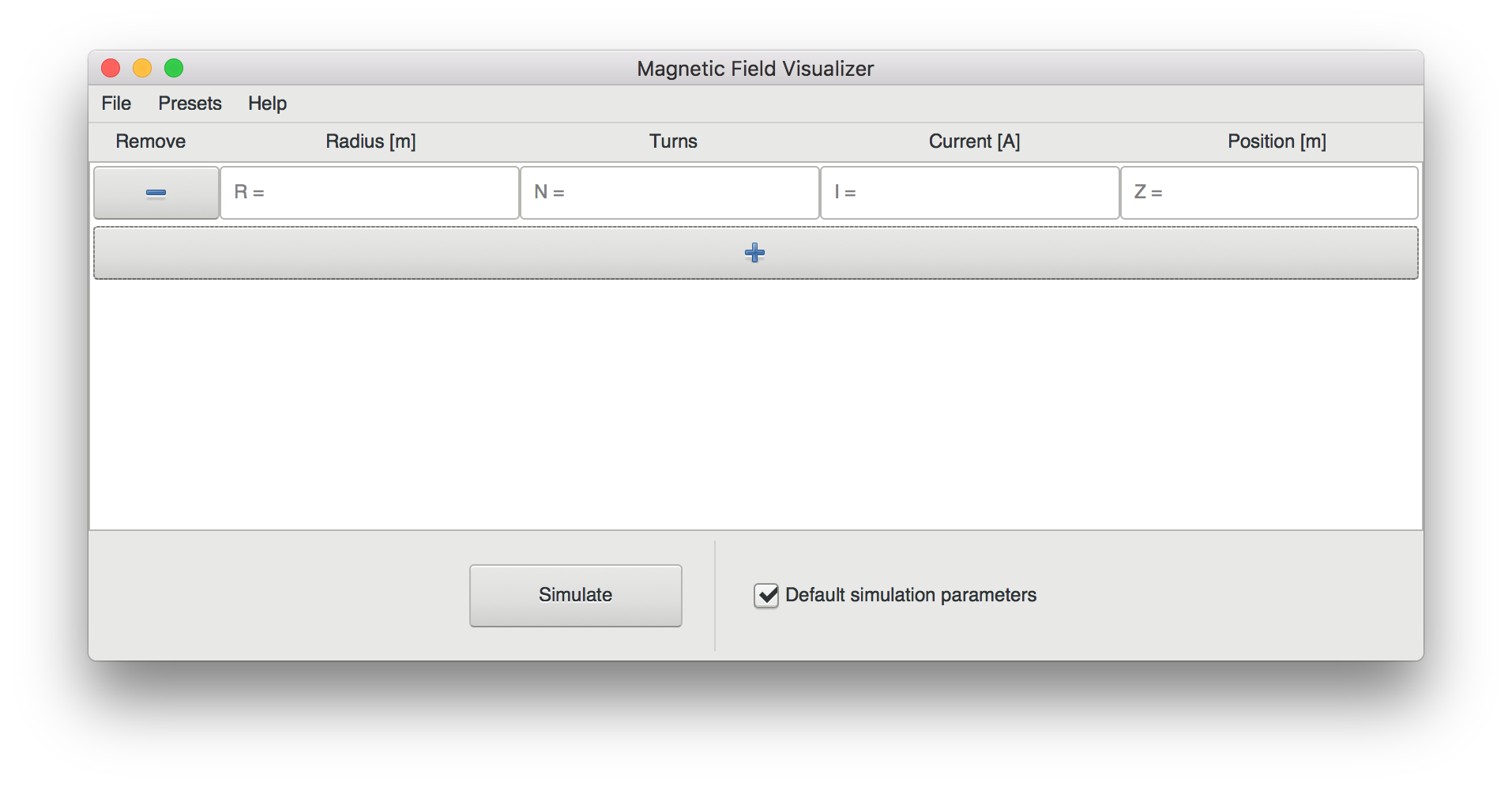}
\par\end{centering}
}
\par\end{centering}
\begin{centering}
\subfloat[\label{fig:simulation-parameters}]{\begin{centering}
\includegraphics[width=0.3\paperwidth]{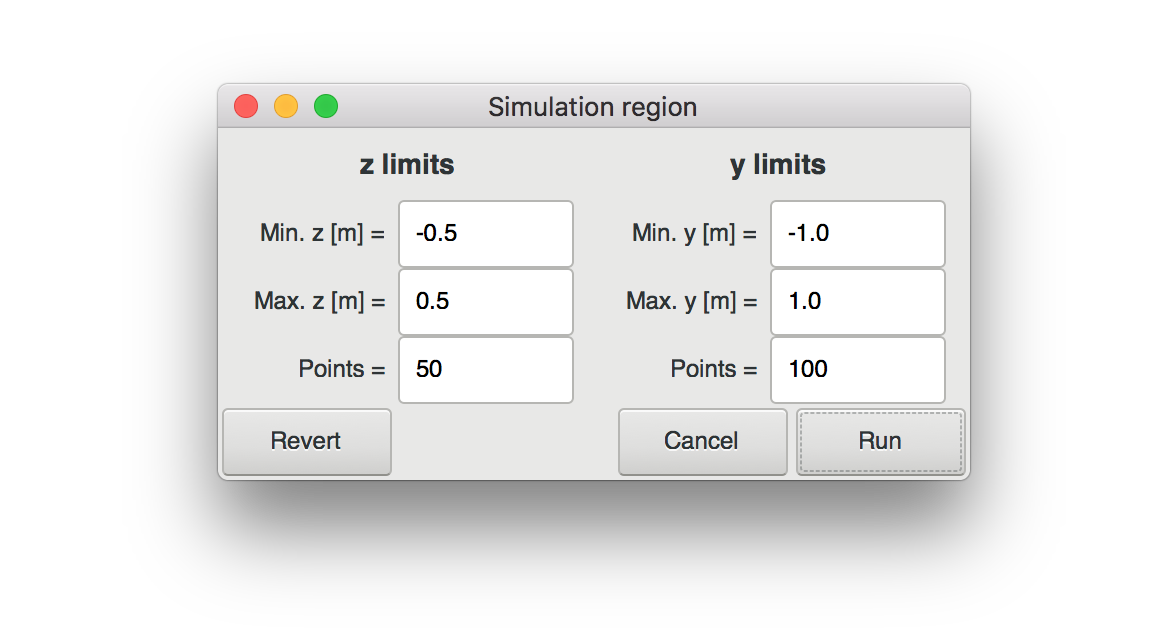}
\par\end{centering}
}
\par\end{centering}
\caption{MFV (a) input and (b) simulation parameters windows.}
\end{figure}
\par\end{center}

\subsubsection{Results}

The results window shows the visualization of the magnetic flux density
distribution, as shown in Figure \ref{fig:results-window}. In this
window, the menu bar includes the ``Colormap'' menu, which contains
a list of different colormaps that can be used for the visualization
of the magnetic field distribution. Also, the ``File'' menu includes
an option to save (``Save as...'') the input parameters and simulation
results in a XLSX file format. The main part of the results window
is composed by three tabs: the first for the visualization of the
magnetic field distribution (see Figure \ref{fig:magnetic-field-distribution-tab}),
the second for the input parameters, and the third for the electric
parameters (see Figure \ref{fig:electric-parameters-tab}) . The graph
of the magnetic flux density distribution is accompanied by a color
bar that indicates the values of the magnetic field in millitesla
($\text{mT}$). The limits of the color bar can be set by using the
``Bmin'' and ``Bmax'' entries located at the lower part of the
window and clicking the ``Apply'' button. Three buttons are also
provided to allow the user to save the graph, reset the color bar
limits, and show or hide the coils in the graph. Graphs are saved
in PDF format by default; however, the user can also save the them
in PNG format. If the user clicks a point in the graph, a message
indicating the coordinates of the point and the value of the magnetic
field at that point appears in the lower part of the window, as shown
in Figure \ref{fig:magnetic-field-distribution-tab}. The electrical
parameters tab gives information about the characteristics of the
wire that should be used to build the simulated coil system based
on the American wire gauge (AWG). 
\begin{center}
\begin{figure}[h]
\begin{centering}
\subfloat[\label{fig:magnetic-field-distribution-tab}]{\includegraphics[width=0.5\columnwidth]{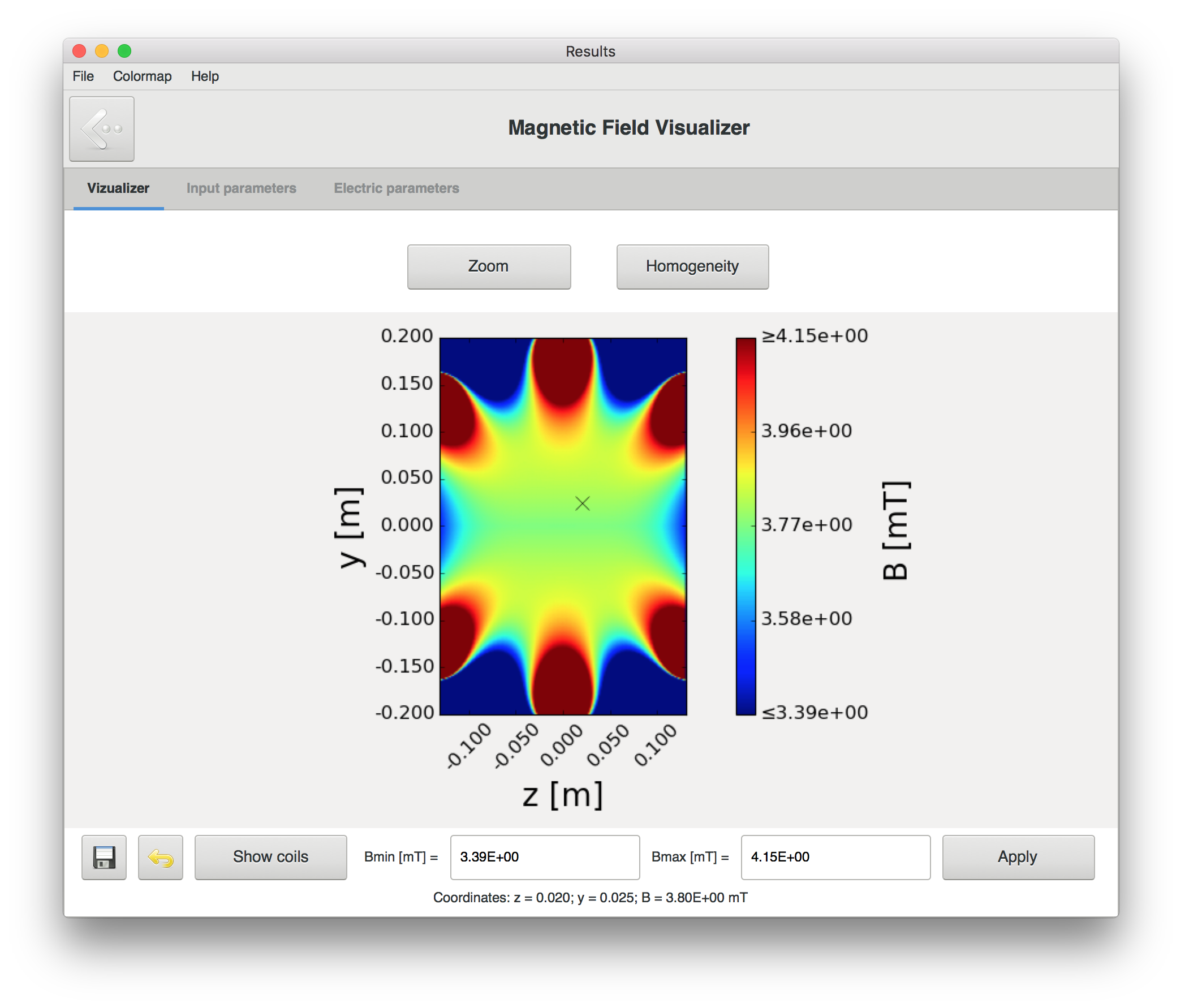}

}\subfloat[\label{fig:electric-parameters-tab}]{\includegraphics[width=0.41\paperwidth]{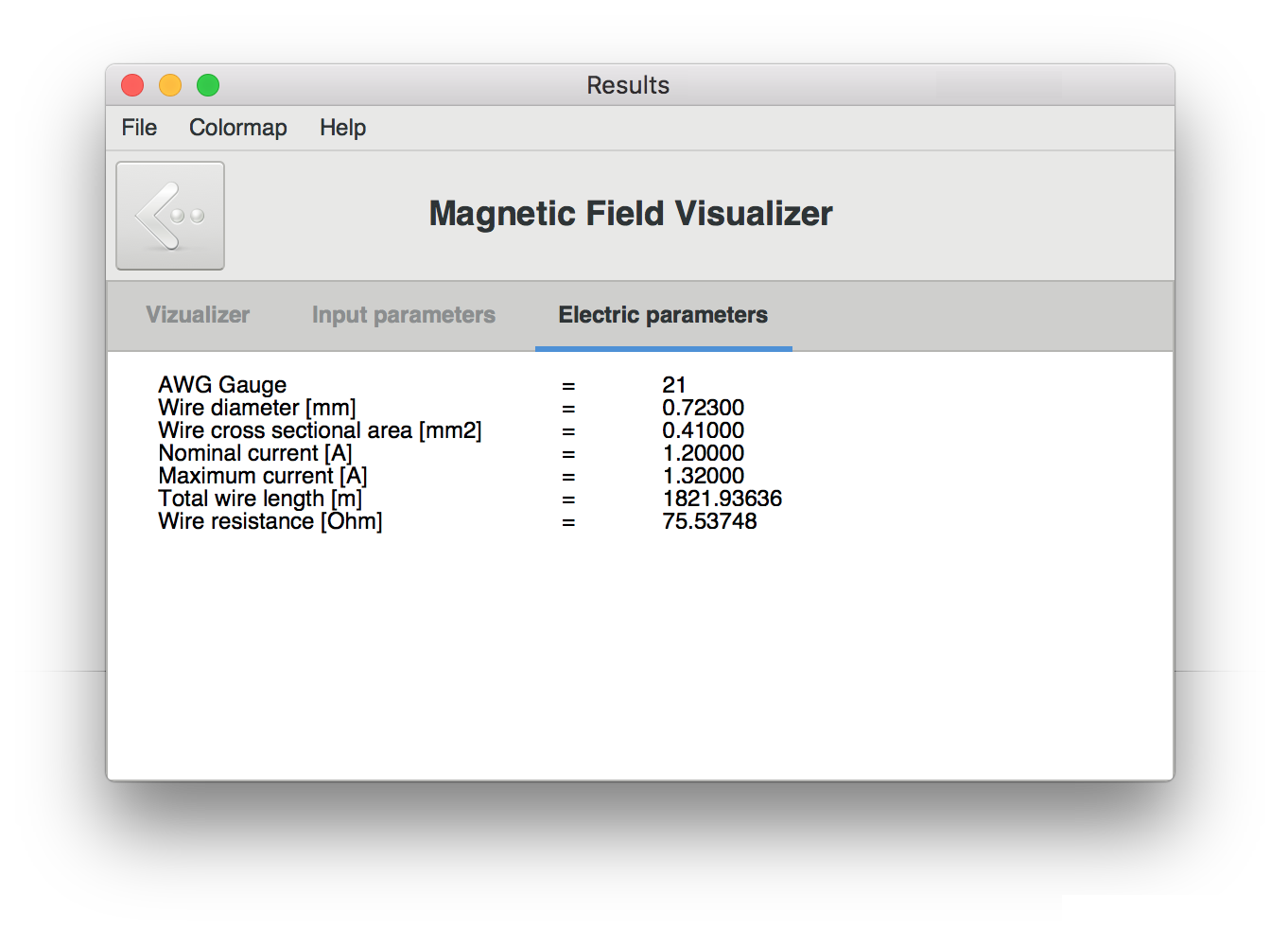}

}
\par\end{centering}
\caption{(a) Magnetic field distribution and (c) electrical parameters tabs
of the results window. The simulated coil system corresponds to a
Maxwell coil.\label{fig:results-window}}
\end{figure}
\par\end{center}

\subsubsection{Zoom and homogeneity}

The two buttons located below the tab bar in the results window, ``Zoom''
and ``Homogeneity'', are used for the zoom and homogeneity functionalities
of the software. When one of these buttons is clicked, a new window
pops up. The zoom functionality allows the user to enter a zoom percentage
to zoom in or zoom out the graph of the magnetic field distribution,
as shown in Figure \ref{fig:zoom-interface}. The homogeneity functionality
takes as input an homogeneity percentage and plots the homogeneous
region according to that percentage, as shown in Figure \ref{fig:homogeneity-interface}.
Because the homogeneous regions are not practical to design an experiment
due to their irregular shape, the largest square that can be inscribed
in the homogeneous region is computed and shown in the graph. The
homogeneity window includes a tab (``Experimentation volume'') where
the characteristics of a practical experimentation volume, produced
due to the rotational symmetry of the circular coil systems, are provided.
The practical experimentation volume corresponds to the solid of revolution
generated by rotating the square region around the axis of symmetry.
The size of the practical experimentation volume and its location
can be useful to the user to know the best position to place a sample
in an experiment. Several zoom and homogeneity windows can be opened
at the same time allowing the user to compare the different graphics.
\begin{center}
\begin{figure}[h]
\begin{centering}
\subfloat[\label{fig:zoom-interface}]{\includegraphics[width=0.48\columnwidth]{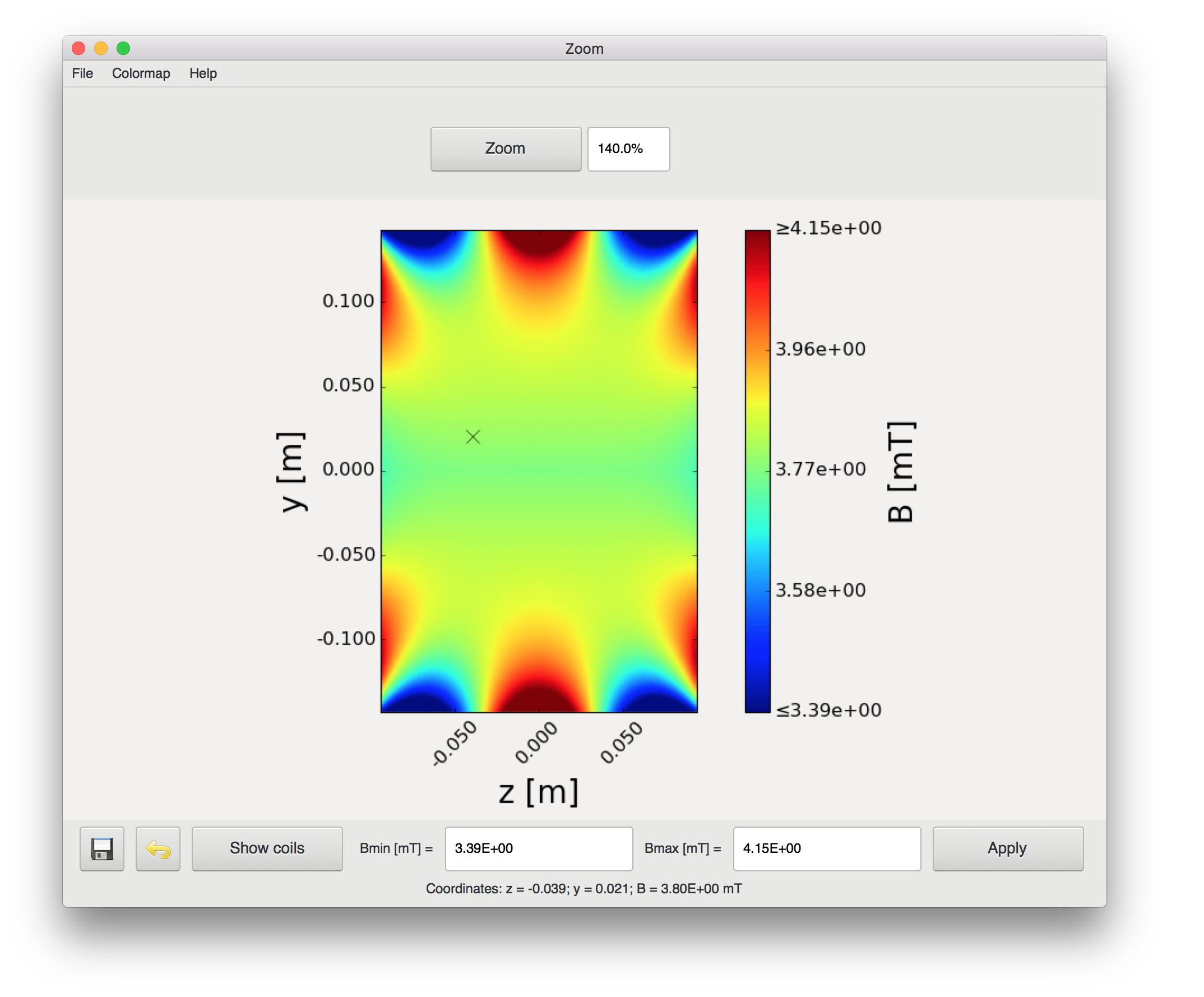}

}\subfloat[\label{fig:homogeneity-interface}]{\includegraphics[width=0.4\paperwidth]{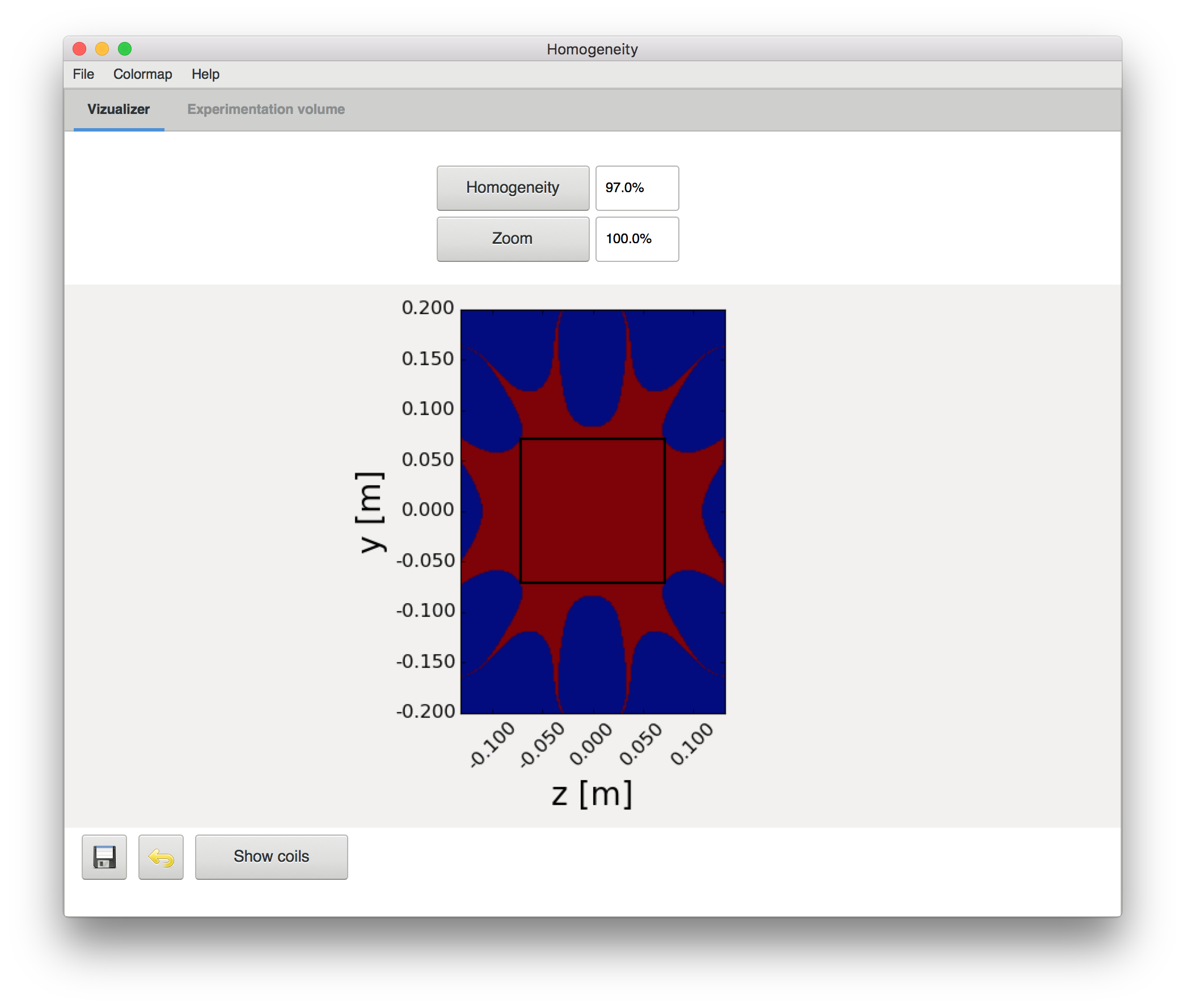}

}
\par\end{centering}
\caption{(a) Zoom and (c) homogeneity windows.}
\end{figure}
\par\end{center}

\section{Applications}

In this section, we present some graphs and data obtained using MFV.
Figure \ref{fig:Magnetic-field-distribution-presets} shows graphs
of the magnetic field distribution of the different presets included
in the software. These graphs show how these coil systems present
different magnetic field behaviors according to the number of coils,
and its distribution and characteristics. Moreover, Figure \ref{fig:homogeneity-presets}
shows the homogeneous regions of the presets for a $97\%$ homogeneity
value. These kind of results give valuable information about the coil
systems, improving the understanding of the magnetic field behavior
and the potential use of a given coil system for a specific application.
\begin{center}
\begin{figure}[h]
\begin{centering}
\subfloat[]{\includegraphics[height=0.25\paperwidth]{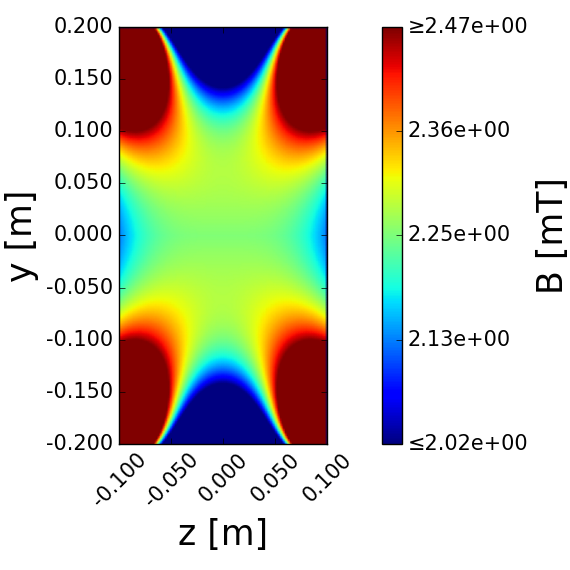}

}\subfloat[]{\includegraphics[height=0.25\paperwidth]{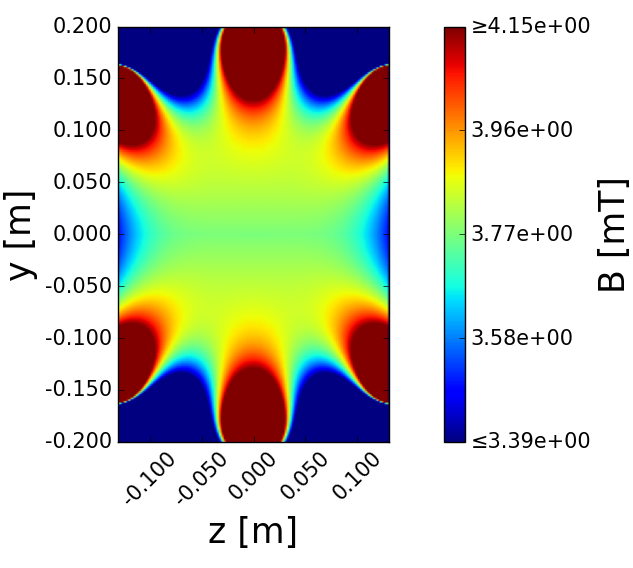}

}\subfloat[]{\includegraphics[height=0.25\paperwidth]{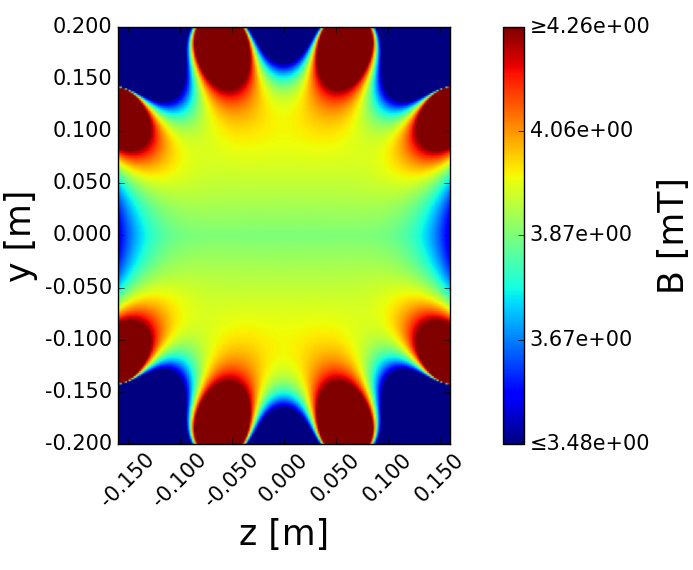}

}
\par\end{centering}
\begin{centering}
\subfloat[]{\includegraphics[height=0.25\paperwidth]{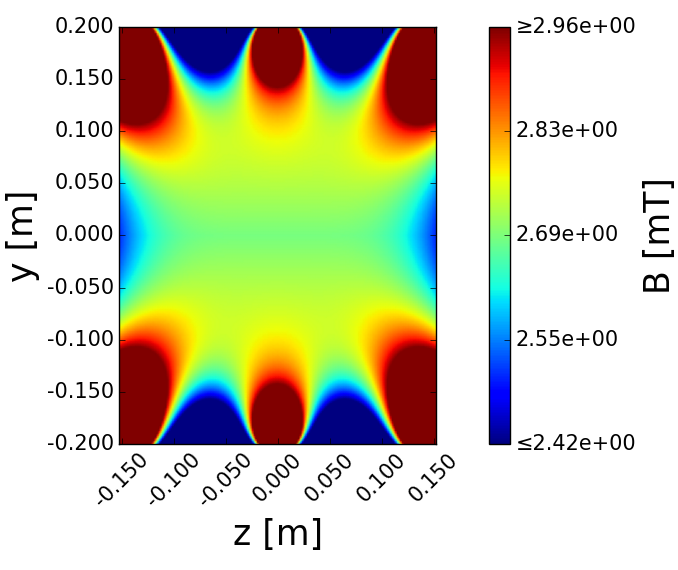}

}\subfloat[]{\includegraphics[height=0.25\paperwidth]{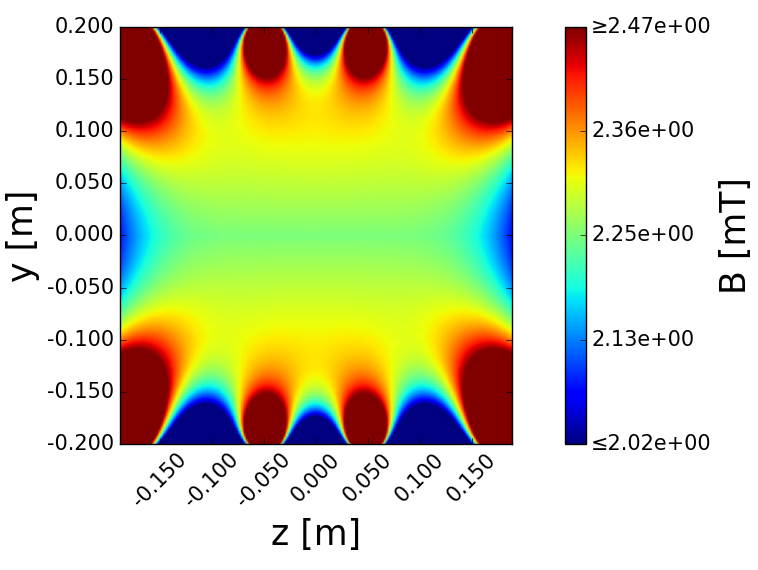}

}
\par\end{centering}
\caption{Magnetic field distribution of (a) Helmholtz coil, (b) Maxwell coil,
(c) Tetracoil, (d) Wang coil, and (e) Lee-Whiting coil systems.\label{fig:Magnetic-field-distribution-presets}}
\end{figure}
\par\end{center}

\begin{center}
\begin{figure}[h]
\begin{centering}
\subfloat[\label{fig:homogeneous-region-a}]{\includegraphics[height=0.25\paperwidth]{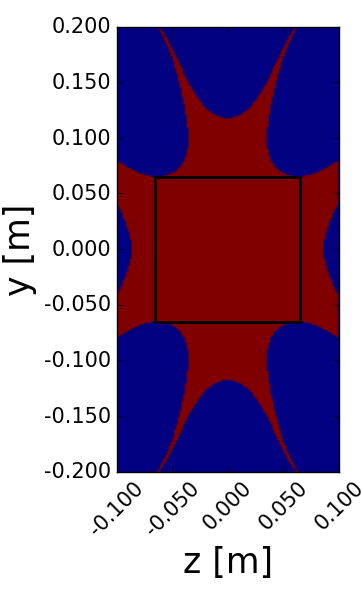}

}\subfloat[]{\includegraphics[height=0.25\paperwidth]{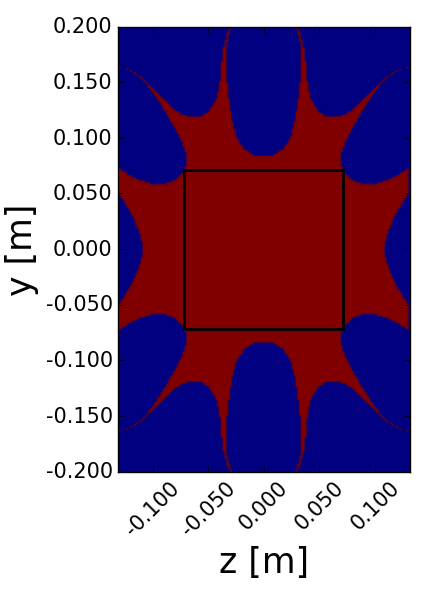}

}\subfloat[]{\includegraphics[height=0.25\paperwidth]{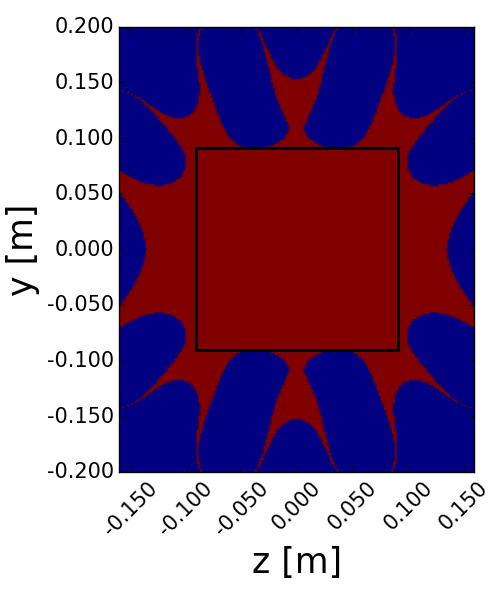}

}
\par\end{centering}
\begin{centering}
\subfloat[]{\includegraphics[height=0.25\paperwidth]{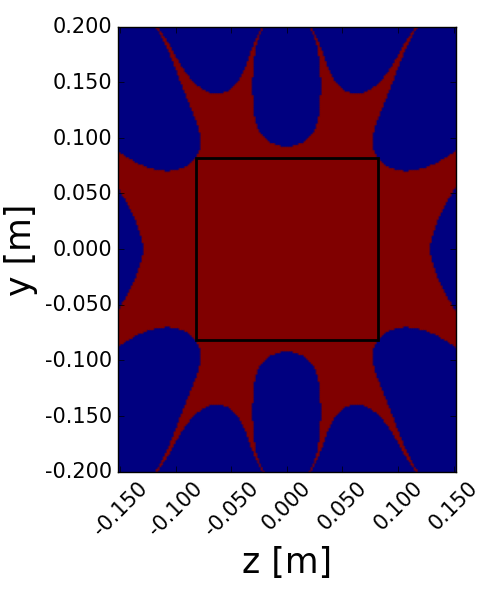}

}\subfloat[]{\includegraphics[height=0.25\paperwidth]{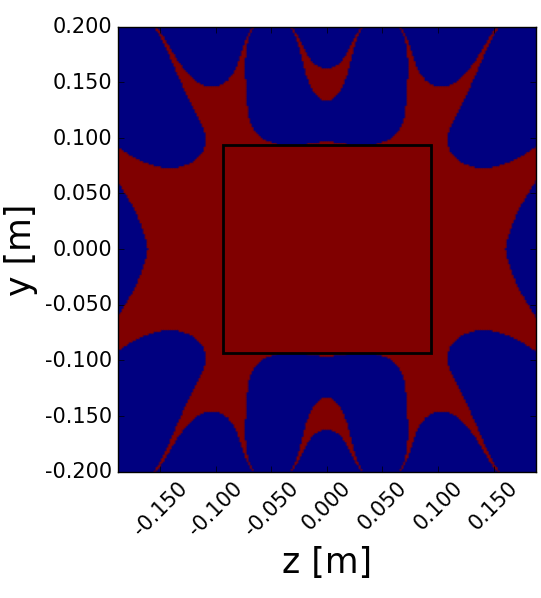}

}
\par\end{centering}
\caption{Homogeneous region of (a) Helmholtz coil, (b) Maxwell coil, (c) Tetracoil,
(d) Wang coil, and (e) Lee-Whiting coil systems for a $97\%$ homogeneity
value.\label{fig:homogeneity-presets}}
\end{figure}
\par\end{center}

As mentioned before, the ``Experimentation volume'' tab in the Homogeneity
window provides useful information of a potential experimentation
volume. Figure \ref{fig:experimentation-volume-params} shows the
experimentation volume parameters of the homogeneous region plotted
in Figure \ref{fig:homogeneous-region-a} for a Helmholtz coil. Because
the square region is located at the center of the Helmholtz coil and
due to the rotational symmetry of the system, the experimentation
volume corresponds to a cylinder. The cylinder has a height, a radius,
and a volume of $0.13029\,\text{m}$, $0.06514\,\text{m}$, and $0.00174\,\text{m}^{3}$,
respectively, and its centroid is located at the center of the system
($z=0.0,y=0.0$). Depending on the location of the homogeneous region
with respect of the axis on symmetry, different experimentation volume
shapes can be produced. The XLSX file, created when the simulation
results are saved, includes the input, electrical, and simulation
parameters, and the $\rho$ component, $z$ component and magnitude
of the magnetic field at each point. From these results, different
analysis of the magnetic field distribution can be made. For instance,
from the results of the magnitude of the magnetic field of the Helmholtz
coil, it is possible to plot the value of the magnetic field ($B$)
along the axis of symmetry ($z$), as shown in Figure \ref{fig:data_plot}.
Moreover, if a new coil system design is simulated, the electrical
parameters provided by the software can be helpful for the building
of the coil system.
\begin{center}
\begin{figure}[h]
\begin{centering}
\subfloat[\label{fig:experimentation-volume-params}]{\includegraphics[width=0.4\columnwidth]{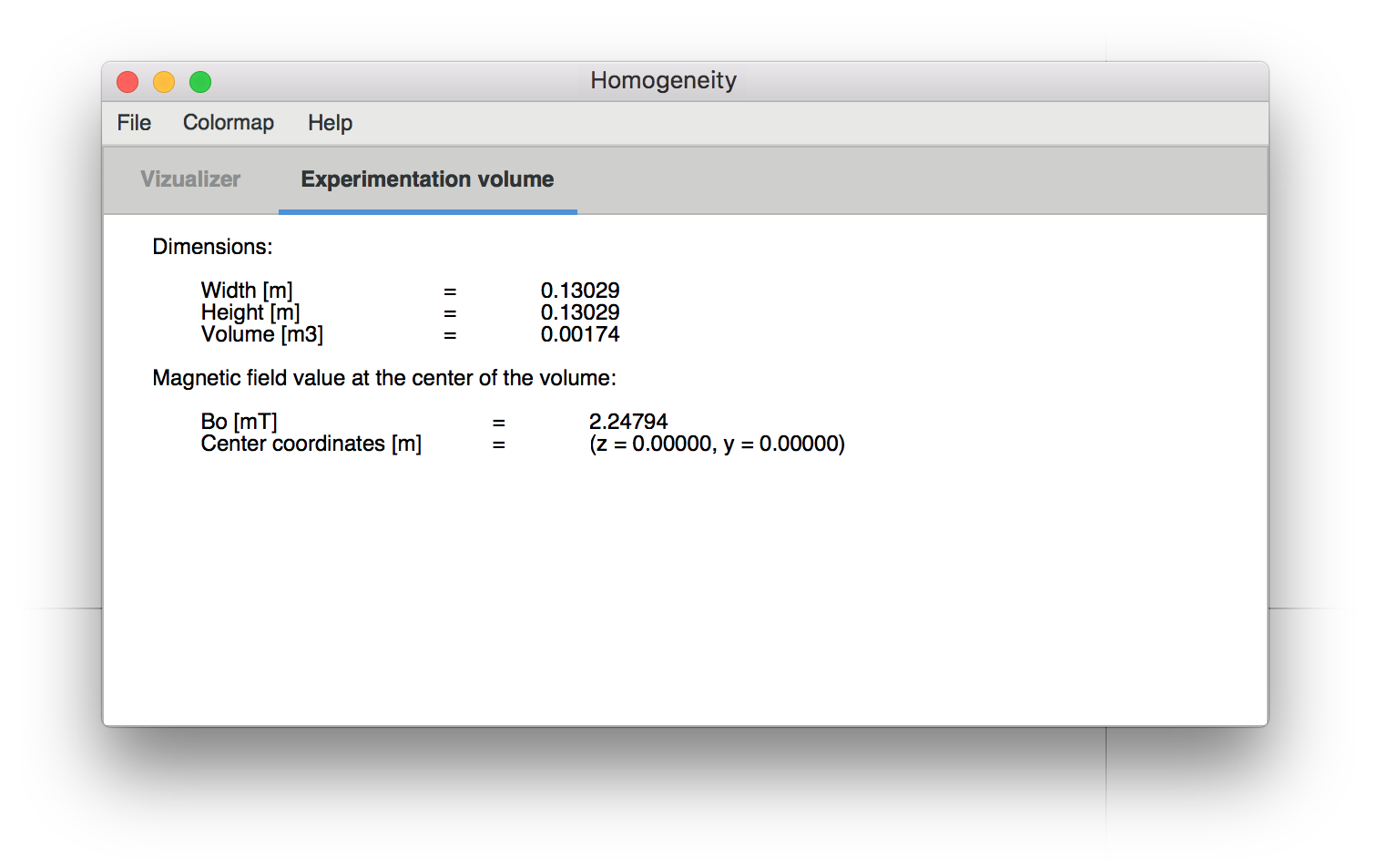}

}\subfloat[\label{fig:data_plot}]{\includegraphics[width=0.4\columnwidth]{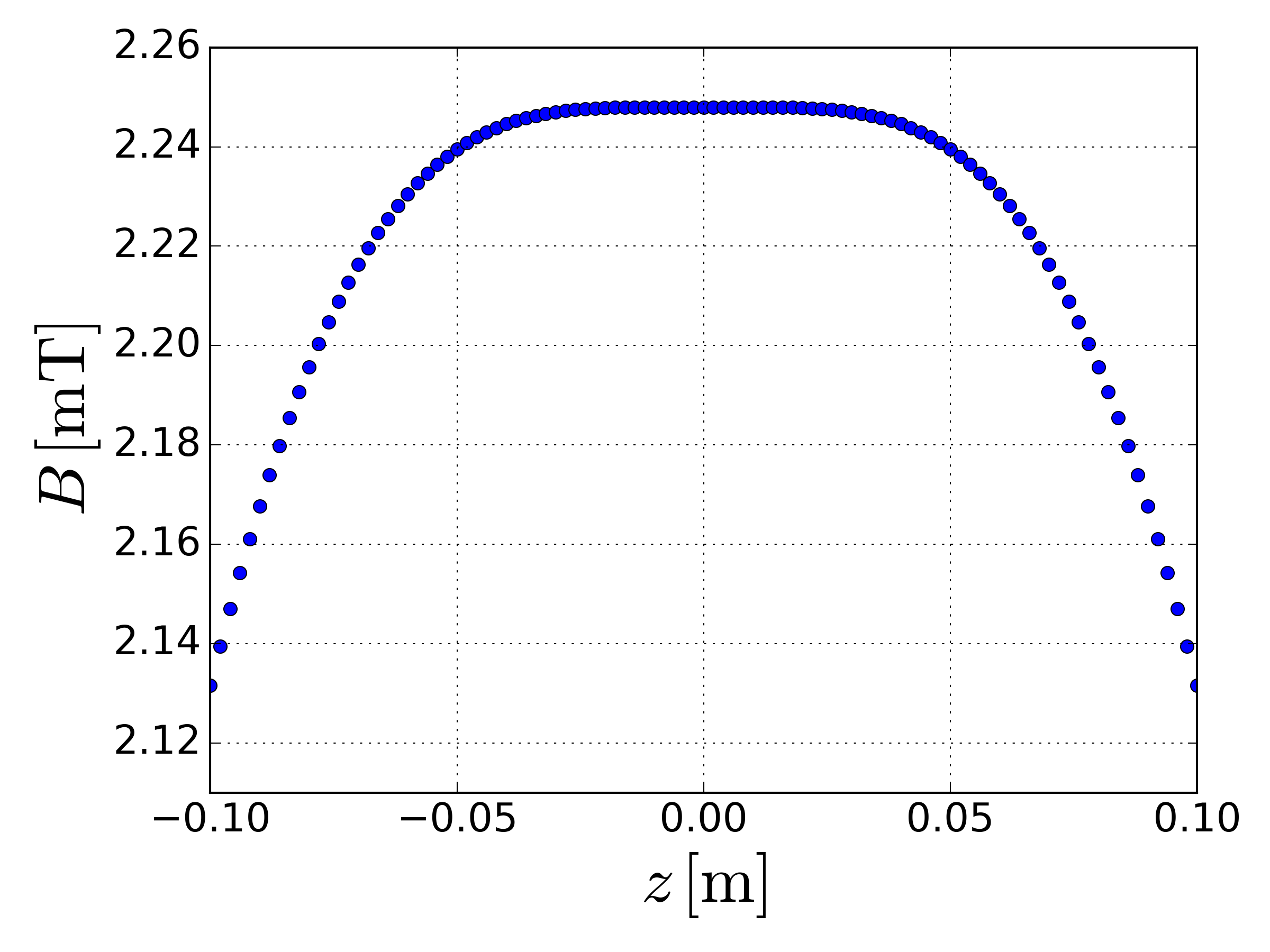}

}
\par\end{centering}
\caption{(a) Experimentation volume parameters and (b) magnitude of the magnetic
field along the axis of symmetry of a Helmholtz coil modeled and simulated
in MFV. }
\end{figure}
\par\end{center}

\section{Perspectives}

Different improvements are planned for MFV. Addition of new coil shapes,
such as circular or rectangular shapes, would allow the modeling and
simulation of other known coil systems (e.g., the Merritt coil \citep{Merritt1983})
and new designs that use different coil shapes. A 3D visualization
of the magnetic field could be useful to give a more complete picture
of the magnetic field distribution in a coil system. The practical
experimentation volume can be optimized by finding the largest rectangle
that can be inscribed inside the homogeneous region; however, that
could require significant computational time and the implementation
of an advanced algorithm. Software developers are encouraged to contribute
to the development of \noun{MFV} at its GitHub repository \citep{Alzate-Cardona2019}.

\bibliographystyle{ieeetr}
\bibliography{bib}

\end{document}